\documentclass[a4paper,UKenglish,cleveref, autoref, thm-restate,numberwithinsect,nameinlink]{lipics-v2021}

\title{Token Sliding Reconfiguration on DAGs} 

\author{Jona Dirks}{University Clermont Auvergne, France}{jona.dirks@uca.fr}{https://orcid.org/0009-0002-9580-9461}{}

\newtheorem{rerule}[theorem]{Rule}

\author{Alexandre Vigny}{University Clermont Auvergne, France}{alexandre.vigny@uca.fr}{https://orcid.org/0000-0002-4298-8876}{}
 
\authorrunning{J. Dirks and A. Vigny} 

\Copyright{Jona Dirks and Alexandre Vigny} 



\category{} 

\relatedversion{} 

\nolinenumbers 

\EventEditors{John Q. Open and Joan R. Access}
\EventNoEds{2}
\EventLongTitle{42nd Conference on Very Important Topics (CVIT 2016)}
\EventShortTitle{CVIT 2016}
\EventAcronym{CVIT}
\EventYear{2016}
\EventDate{December 24--27, 2016}
\EventLocation{Little Whinging, United Kingdom}
\EventLogo{}
\SeriesVolume{42}
\ArticleNo{XX}

\usepackage{booktabs}

\usepackage[]{blindtext}
\usepackage{complexity}
\usepackage{algorithm2e}
\usepackage{subcaption}
\usepackage{subfiles}
\usepackage{wasysym}
\usepackage{lineno} 

\usepackage{amsthm}
\usepackage{mathrsfs}
\usepackage{mathtools}
\usepackage{todonotes}
\usepackage{float}

\newcommand{\tw}{\ensuremath{\textrm{tw}}}

\usepackage{tikz}
\usetikzlibrary{decorations.pathmorphing, calc}
\usetikzlibrary{arrows.meta}
\pgfdeclarelayer{background}
\pgfdeclarelayer{foreground}
\pgfsetlayers{background,main,foreground}

\newcommand{\MSO}{\ensuremath{\mathsf{MSO}}}

\newcommand{\bag}{\mathsf{bag}}
\newcommand{\cone}{\mathsf{cone}}
\newcommand{\cmp}{\mathsf{comp}}
\newcommand{\mrg}{\mathsf{mrg}}
\newcommand{\adh}{\mathsf{adh}}

\newcommand{\warp}{\mathsf{warp}}
\newcommand{\parent}{\mathsf{parent}}

\newcommand{\significant}{\mathsf{significant}}
\newcommand{\hull}{\mathsf{hull}}
\newcommand{\rightsignificant}{\mathsf{right\textrm{-}significant}}

\newcommand{\isdts}{ISR-DTS\xspace}
\newcommand{\ists}{ISR-TS\xspace}

\newcommand{\istj}{ISR-TJ\xspace}
\newcommand{\ts}{TS\xspace}
\newcommand{\tj}{TJ\xspace}

\usepackage{contour}
\usepackage{ulem}

\contourlength{0.8pt}

\renewcommand{\underline}[1]{%
	\uline{\phantom{#1}}%
	\llap{\contour{white}{#1}}%
}

\begin{document}
    \tikzset{vertex/.style={circle, color=black, draw=black, minimum size=0.7cm}}
    
    \maketitle

    
    \begin{abstract}
    	Given a graph $G$ and two independent sets of same size, the {\em Independent Set Reconfiguration Problem under token sliding} asks whether one can, in a step by step manner, transform the first independent set into the second one. In each step we must preserve the condition of independence. Further, referring to solution vertices as tokens, we are only permitted to slide a token along an edge.
    	Until the recent work of Ito et al.~[Ito et al. MFCS 2022] this problem was only considered on undirected graphs. In this work, we study reconfiguration under token sliding focusing on DAGs. 
    	
    	We present a complete dichotomy of intractability in regard to the depth of the DAG, by proving that this problem is \NP-complete for DAGs of depth 3 and $\W[1]$-hard for depth 4 when parameterized by the number of tokens $k$, and that these bounds are tight.	Further, we prove that it is fixed parameter tractable on DAGs parameterized by the combination of treewidth and $k$. We show that this result applies to undirected graphs, when the number of times a token can visit a vertex is restricted.
    \end{abstract}
	
	\newpage
	\setcounter{page}{1}
	\section{Introduction}

A reconfiguration problem asks, given two solutions to a combinatorial problem, whether one can change the first solution in a step-by-step manner in order to reach the second one; while ensuring that the transformed set remains a solution throughout the reconfiguration procedure.
The research interest in these topics dates back to the 19th century, as many one player games can be considered in these terms~\cite{Johnson1879}. For a more recent introduction to the topic, we refer to~\cite{Nishimura18}.

The complexity of reconfiguration problems is often \PSPACE-complete. Specific attention has been given to independent set and (connected) dominating set reconfiguration. For these problems, there are two main reconfiguration paradigms: {\em token sliding}, and {\em token jumping}. In both paradigms, we view the set to be reconfigured as several tokens placed on the vertices of the graph. Token jumping allows, in each step to move one token to be placed to any other vertex of the graph, while token sliding only allows to move a token to an adjacent vertex (to slide along an edge of the graph).

For both paradigms, independent set reconfiguration is \PSPACE-complete, and remains intractable when parameterized by the size of the set to be reconfigured, or the length of reconfiguration sequence~\cite{BodlaenderGS21,MouawadN0SS17}.
In order to find efficient algorithms, it is therefore interesting to look at structural restrictions on the graphs over which the sets are being reconfigured.
While these problems remain \PSPACE-hard on planar graphs and graphs of bounded bandwidth, they become polynomial on cographs, forest and interval graphs~\cite{HaddadanIMNOST16,LokshtanovM19,Wrochna2018}.
We however note that the two paradigms differ in complexity over bipartite graphs, where independent set reconfiguration for token sliding (\ists) is \PSPACE-complete, independent set reconfiguration for token jumping (\istj) is only \NP-complete~\cite{LokshtanovM19}. 

The two paradigms differ even more when we look at parameterized complexity. For token jumping, we know that reconfiguration for independent set over general sparse graphs classes, such as bounded degeneracy and nowhere dense, becomes \FPT\xspace when parameterized by the size of the sets~\cite{Bousquet2024,LokshtanovM19,Siebertz18}. These classes are generalizations of many well known classes such as planar graph, graph of bounded treewidth, classes excluding a (topological) minor etc.

For token sliding however, the parameterized complexity of independent set reconfiguration is much less understood.
On one hand it is known to be \FPT\xspace on trees~\cite{DemaineDFHIOOUY14}, graphs of bounded degree~\cite{Bartier2023}, and graphs of girth 5~\cite{BartierBHMS24}, but hard on split graphs~\cite{BonamyB17}.
On the other hand, it is still open whether it is tractable on graph of bounded treewidth, or minor-free graphs~\cite[Question 5]{Bousquet2024}. 
We refer the readers to the survey~\cite{Bousquet2024} for an extensive overview.

We also mention that for both \ts and \tj the reconfiguration of every problem that is expressible in $\MSO_2$ is \FPT\xspace when parameterized by treedepth and the number of tokens~\cite[Theorem 4.1]{Gima2024}
and \FPT\xspace when parameterized by treewidth, the length of the reconfiguration sequence, and the formula describing the considered problem~\cite[Theorem 7]{MouawadNRW14}

\medskip\noindent
{\bf Directed graphs.}
In this paper, we consider directed graphs and allow tokens to only slide in the direction of the edges. Note, that token jumping is not well suited for directed graphs, as arbitrary jumps stay the same in the directed setting. So far very little is known for directed graph's reconfiguration.
Recently, Ito et al.~\cite[Theorem 2]{ito2022} proved that Independent Set Reconfiguration for Directed Token Sliding (\isdts) is \NP-hard on DAGs and \W[1]-hard when parameterized by the number of tokens.
On the tractable side of things, they show that it is linear time solvable on orientation of trees~\cite[Theorem 3]{ito2022}. The hardness extends to \PSPACE-completeness for oriented planar graphs, split graphs, bipartite graphs and most noticeable bounded treewidh graphs \cite{Banerjee2024}.

We improve on both the side of hardness as well as the side of tractability. We first look at the depth of the DAGs and provide sharp dichotomy in both the classical and parameterized setting. In the following statements, $k$ refers to the number of tokens. 

\begin{restatable}{theorem}{tractability}
	\label{theo:tractability}
	\isdts is polynomial on DAGs of depth~2 and \FPT\xspace for DAGs of depth~3 when parameterized by  $k$. 
\end{restatable}

\begin{restatable}{theorem}{npcompletedepthtwo}
	\label{theo:np-complete-depth-2}
	\isdts is \NP-complete on DAGs of depth 3. 
\end{restatable}

\begin{restatable}{theorem}{woneharddepthfour}
	\label{theo:w1-hard-depth-4}
	\isdts is $W[1]$-hard on DAGs of depth 4, when parameterized by $k$.
\end{restatable}

Additionally, we show that a combination of the parameters $k$ and treewidth leads to tractability in DAGs. We do so by adapting the recently developed framework of galactic reconfiguration \cite{Bartier2023} to directed graphs. The following result can be considered our main contribution.
\begin{restatable}{theorem}{fpttwk}
	\label{theo:fpt-tw-k}
	\isdts is \FPT\xspace on DAGs when parameterized by $k+\tw$.
\end{restatable}

Finally, we go back to undirected graph, introduce a new parameter (the iteration $\iota$) that measures the maximum number of times a specific token can enter a specific vertex. 
\begin{restatable}{theorem}{fptiota}\label{theo:fpt-iota}
	\ists is \FPT\xspace when parameterized by $k+\tw+\iota$.
\end{restatable}

These results for graphs of bounded treewidth is a step toward solving one of the most prominent open problem in this line of work (see~\cite[Question 5]{Bousquet2024}): is \ists\xspace \FPT\xspace when parameterized by $k$ and the treewidth of the graph?

The previous best algorithms only work on the more restricted class of bounded treedepth~\cite{Gima2024}, or require to also parameterize by the length $\ell$ of the reconfiguration sequence~\cite{MouawadNRW14}.
In comparison, our new parameter {\em iteration} is much smaller than the length of the reconfiguration sequence. In terms of techniques, parameterizing by $\ell$ enabled the authors of~\cite{MouawadNRW14} to reduce to the model checking problem for \MSO\xspace formula which is solved using Courcelle's Theorem. In our case, we do not believe that we can reduce to Courcelle's theorem. Thus, the presented algorithm is the first to use dynamic programming on graph of bounded treewidth ``by hand'' for Independent Set Reconfiguration.

\medskip
\noindent{\bf Organization.}
The rest of the paper is structured as follows. \cref{sec:prelim} provides the required background and fixes notations. \cref{sec:tractability} is devoted to efficient algorithms and the proof of \cref{theo:tractability}. The 
proof of \cref{theo:np-complete-depth-2} is provided in \cref{sec:np-hard}, followed by the proof of \cref{theo:w1-hard-depth-4} in \cref{sec:w1-hard}. In \cref{sec:tw} we introduce additional notions around galactic reconfiguration for our main algorithm in order to prove \cref{theo:fpt-tw-k}. Finally, in \cref{sec:undirected} we discuss implications for undirected graphs, and \cref{theo:fpt-iota}.

	\section{Preliminaries}\label{sec:prelim}
	\medskip
\noindent{\bf Graphs.}
We use standard graph notations. All graphs are directed and loop-free unless stated otherwise. Unless preferable for readability we abbreviate an edge $(u,v)$ with $uv$. We denote the open out-neighborhood of a vertex $v$ with $N^+(v)$ and the open in-neighborhood with $N^-(v)$. A graph without cycles is called a DAG. The {\em depth} of a DAG is the number of vertices on its longest directed path. Note that if a DAG has depth $d$, then its vertices can be grouped in $d$ levels with every edge going from a vertex at some level to a vertex at a greater level. One can inductively define the first layer as the vertices of in-degree~0, and the $i$th layer as the vertices whose entire in-neighborhood is in smaller levels.

For a directed graph $G$, the underling undirected graph is obtained by disregarding the edge direction and removing all double edges. That is the graph $(V(G), \{\{u,v\} \mid uv \in E(G) \textrm{ or } vu\in E\})$. A set of vertices of a directed or undirected graph is called {\em independent} if the vertices are pairwise nonadjacent.

\medskip
\noindent{\bf Parameterized Complexity.}
In the parameterized framework, a problem is said to be {\em fixed parameter tractable} (\FPT), if there exists an algorithm that solves it in time $f(k)\cdot n^c$ for a computable function $f$ and constant~$c$.
An alternative characterization of \FPT\xspace is through the existence of a \emph{kernel}, that is polynomial time computable function that maps each instance to an instance of size $g(k)$ for some computable function $g$. The returned instance should be a positive-instance exactly if the inputted instance is. Kernels are commonly presented in the form of data reduction rules, that provide instructions on how to shrink the size of the input.

Under commonly made assumptions, problems that are $\W[1]$-hard are considered not to be \FPT. Similar to \NP-completeness, we use \FPT-reductions to transfer hardness. However, these reductions may use a longer \FPT\xspace running time but are more restricted in the sense that the parameter of the resulting instance may not depend on the size but only on the parameter of the inputted instance.
More background on \FPT\xspace algorithms and \FPT-reduction can be found in the book~\cite{book15parameterized-algorithms}.

\medskip
\noindent{\bf Reconfiguration.}
A {\em configuration} $\alpha_i$ is a function from a set of tokens $T=\{t_1,t_2,\dots, t_k\}$ to the set of vertices $V$.

For two independent sets $S$ and $D$ of size $k$ in a graph $G$, a \emph{reconfiguration-sequence} for \isdts is a sequence of configurations $(\alpha_i)_{i\in I}$ such that:
\begin{enumerate}
	\item $\alpha_1(T)=S$ and $\alpha_{|I|}(T) = D$.
	\item For every $i\in I$, and for distinct tokens $t$ and $t'$: $\alpha_i(t) \neq \alpha_i(t')$.
	\item For every $i\in I$, and for distinct tokens $t$ and $t'$: $(\alpha_i(t),\alpha_i(t'))\not\in E(G)$.
	\item If $\alpha_{i}(t) \neq \alpha_{i+1}(t)$, then there exists an edge $(\alpha_{i}(t),\alpha_{i+1}(t)) \in E(G)$ and for every other token $t'\neq t$ it holds that $\alpha_{i}(t') = \alpha_{i+1}(t')$.
\end{enumerate}

We generally assume that $I$ is a sequence from 1 to $|I|$. For a configuration $\alpha_i$ we call all configurations $\alpha_{j}$ \emph{succeeding} or a \emph{successor}, if $j>i$. If~$j=i+1$ we say that it is a \emph{direct} successor.

Let $t$ be a token and $\alpha=(\alpha_i)_{i\in I}$ a sequence with $\alpha_1(t) = u$ and $\alpha_{|I|}(t) = v$. We say that the token $t$ \emph{comes from} $u$ and \emph{ends up in} $v$. 
Similarly we say that a token \emph{comes from} a subgraph $A$ and \emph{ends up in} a subgraph $B$ if $\alpha_1(t)\in V(A)$ and $\alpha_{|I|}(t)\in V(B)$. Additionally, if there is a token on $v$ we say that this token \emph{blocks} all $u\in N(v)$.

\medskip
\noindent {\bf Galactic Graphs.}
Galactic graphs were recently developed to prove that \ists is in \FPT\xspace for undirected graphs of bounded degree~\cite{Bartier2023}.
We restate their definition, slightly adapted for directed graphs. A \emph{galactic graph} is a graph where the vertex set is partitioned $V(G) = A(G) \cup B(G)$ into planets $A(G)$ and black holes $B(G)$.

For a (galactic) graph $G$ and a subsets of its vertices $V'$ we say that it \emph{collapses} to $G'$ if $G-V' = G'-b$ with $b\in B(G')$ and $G'$ contains an edge $(u,b)$ (respectively $(b,u)$) if $G$ contains an edge $(u,v)$ (respectively $(v,u)$) for some $v\in V'$. We denote this graph with~$G\odot V'$. For the empty set, we define $G\odot\emptyset = G$.
Intuitively, this operation takes a set of vertices (planets and black holes) and merges them into one black hole. 

Note, that this operation takes a graph on the left-hand site and a vertex set on the right-hand site. Thus, multiple applications of this operator are left-associative.

We will assume that the black holes are named (or labeled) in such a way that they are uniquely identified by the planets that collapsed into it in the original graph. Meaning that for example for disjoint $V_1, V_2$ it holds $G\odot V_1 \odot V_2 \odot \{b_1, b_2\} = G\odot V_1\cup V_2$, where $b_1$ and $b_2$ are the black holes created when collapsing $V_1$ and $V_2$ respectively. This can be realized by representing planets as sets containing a single element and black holes as the union of the sets representing vertices (planets or black holes) with an additional marker for black holes added.

\medskip
\noindent{\bf Galactic Reconfiguration.}
For two multisets $S$ and $D$ of size $k$ in a graph $G$, a {\em galactic reconfiguration sequence} for \isdts is a sequence of configurations $(\alpha_i)_{i\in I}$. Where $\alpha_i$ is a function from a set of tokens $T=\{t_1,t_2,\dots, t_k\}$ to $V$, such that:
\begin{enumerate}
	\item $\alpha_1(T)=S$ and $\alpha_{|I|}(T) = D$.
	\item For distinct tokens $t$ and $t'$: $\alpha_i(t) \neq \alpha_i(t')$ or $\alpha_i(t) \in B(G)$.
	\item All $\alpha_1(T)\cap A(G),\dots, \alpha_{\ell}(T)\cap A(G)$ are independent sets.
	\item If $\alpha_{i}(t) \neq \alpha_{i+1}(t)$, then there exists an edge $(\alpha_{i}(t),\alpha_{i+1}(t)) \in E(G)$ and for every other token $t'\neq t$ it holds $\alpha_{i}(t') = \alpha_{i+1}(t')$.
\end{enumerate}

Note that in a galactic reconfiguration sequence, several tokens are allowed to be on the same vertex (including for the first and last configuration), as long as this vertex is a black hole. When we collapse an instance for \isdts and the collapsed vertex set contains start and end positions, we generally assume that these positions in their multiplicity get assigned to the created black hole.

\begin{definition}
	Given a graph $G$, a set of vertices $V_1$, we say that a reconfiguration sequence $(\alpha_i)_{i\in I}$, {\em collapses} to $(\beta_j)_{j\in J}$ over $G\odot V_1$ if $(\beta_j)_{j\in J}$ can be obtained by first creating an intermediate reconfiguration sequence $(\beta'_i)_{i\in I}$ where every $\beta'_i$ is mapping to $b$ (the created black hole) every token that $\alpha_i$ maps to a vertex of $V_1$ but is otherwise identical. And then removing exhaustively all configurations with an identical direct successor from $(\beta'_i)_{i\in I}$.

	Moreover, if a sequence $\alpha$ on a graph $G_1$, collapses to a sequence $\beta$ on a graph $G_2$, which itself collapses to a sequence $\gamma$ on a graph $G_3$, we also say that $\alpha$ collapses to $\gamma$ on $G_3$.
\end{definition}
Note that in the above definition, $(\beta_j)_{j\in J}$ is a valid reconfiguration sequence, i.e.~the tokens only slide along edges and are only on neighboring vertices if at least one of them is a black hole.
Furthermore, form this definition, it might look like a single reconfiguration sequence can collapse to multiple sequences on the same graph (depending on the chain of collapses). We later show in \cref{lem:transitivity} that it is not the case, and the collapse notion (with several successive collapses) is therefore well-defined.

\medskip\noindent {\bf Tree-decomposition.}
A \emph{tree decomposition} of a undirected graph $G$ is a pair $\mathcal{T}=(T,\bag)$ where $T$ is a rooted tree and $\bag\colon V(T)\to 2^{V(G)}$ is a mapping that assigns to each node $x\in T$ its bag $\bag(x)\subseteq V(G)$ such that the following conditions are satisfied:
\begin{itemize}
	\item For every vertex $v\in V(G)$, the set of nodes $x\in V(T)$ satisfying $v\in\bag(x)$ induces a connected subtree of $T$.
	\item For every edge $\{u,v\}\in E(G)$, there exists a node $x\in V(T)$ such that $u,v\in\bag(x)$.
\end{itemize}

We may assume that every internal node of $T$ has exactly two descendants. See \cite{book15parameterized-algorithms} for more details on tree decompositions in general. Further, let $\mathcal{T}=(T,\bag)$ be a tree decomposition of a graph $G$.
The \emph{adhesion} of a node $x\in V(T)$ is defined as $\adh(x)\coloneqq \bag(\parent(x))\cap \bag(x)$ and the \emph{margin} of a node $x\in V(T)$ is defined as $\mrg(x)\coloneqq \bag(x)\setminus \adh(x)$.
The \emph{cone} at a node $x\in V(T)$ is defined as $\cone(x)\coloneqq \bigcup_{y\succeq_T x} \bag(y)$ and the \emph{component} at a node $x\in V(T)$ is defined as $\cmp(x)\coloneqq \cone(x)\setminus \adh(x)= \bigcup_{y\succeq_T x} \mrg(y)$.
Here, $y\succeq_T x$ means that $y$ is a descendant of $x$ in $T$. 
The {\em width} of a tree decomposition is the maximum number of vertices in each bag (minus one).
The {\em treewidth} of a graph $G$, noted $\tw(G)$, is the minimum width among all possible tree decompositions of $G$.
For directed graphs, we define these terms over their underlying undirected graphs.
	
	\section{Tractability Results for DAGs of Low Depth}\label{sec:tractability}
	
This section is devoted to the proof of \cref{theo:tractability} that we restate here.
\tractability*

We prove the first part of the theorem. 
\begin{lemma}
	\label{lem:poly-solveable}
	For DAGs of depth 2 \isdts is polynomial solvable.
\end{lemma}
\begin{proof}
	We are given a DAG $G$ and two independent sets $S$ and $D$ both of size $k$.
	First, we can assume that $S\cup D$ is the vertex set of the entire graph, as for every other vertex $v$ no token can start in $S$, go to $v$ and then end in $D$, contradicting that $G$ is a DAG of depth 2.

	If $(G,S,D)$ is a positive instance, there needs to be at least one vertex $u$ in $D$ that has just a single neighbor $v$ in $S$, as otherwise no token can move. We can therefore start the reconfiguration sequence by moving the token from $u$ to $v$. We can now remove $u,v$ from $G$ and proceed inductively in the same way. If at any point no token can move, we report that the input is a negative instance to \isdts. 
\end{proof}

After this small warm up proof, let us now move to the parameterized tractability result. Let us first show, that we can assume the input to have a uniform structure.

\begin{lemma}
	\label{lem:dag-depth-3-regular}
	For any instance $(G, S, D)$ if $G$ is a DAG of depth 3 we can in polynomial time find an equivalent instance $(G', S', D')$ such that $S'$ is equal to the first layer of $G'$ and $D'$ to the third and where $G'$ is also a DAG of depth 3.
\end{lemma}
\begin{proof}
	First note, that if there is a vertex from $S-D$ at the third layer, it can reach nothing and thus we can return a trivial no instance. While a vertex in $S\cap D$ in the third layer cannot move, blocking its in-neighborhood, so we can safely remove this vertex together with it's in-neighborhood. By analog arguments the same holds if there is a vertex from $D$ at the fist layer.
	Additionally, we can remove all vertices without predecessors in $S$ and all vertices with no successors in $D$.
	
	Observe, that after this a vertex from $S$ may not lie on the second layer. If it were so, it would have a predecessor, which then would lie in the first layer. Thus, it would be in $S$ which means that $S$ is not an independent set.
	
	Furthermore, if a vertex $v \in D$ is in the second layer, it may not have successors. Thus, we can add a new vertex $v'$ make it a successor to all predecessors of $v$ as well as $v$ itself. Then, while keeping $v$ in the graph, we remove $v$ from $D$ and add $v'$ instead.
	
	Note that all of these basic checks as well as basic graph manipulations can be carried out in polynomial time.
\end{proof}

We show that the following reduction rule, when applied exhaustively, yields a kernel:

\begin{rerule}
	\label{rule:not-the-same-neighbourhoods}
	If a DAG of depth 3 has two vertices $u,v\notin S \cup D$ with $N(u) = N(v)$ then remove $u$.
\end{rerule}
\begin{proof}[Proof of Correctness]
	Let $G'$ be the graph obtained from applying the reduction rule on~$G$. Clearly if there is a reconfiguration sequence in $G'$ one can use the same sequence in $G$. Thus, we only have to show, that if there is a reconfiguration sequence in $G$ there is one in $G'$.
	
	Let $(\alpha_i)_{i\in I}$ be the reconfiguration. We show that every time $(\alpha_i)_{i\in I}$ places a token on~$u$ we can place it on~$v$ instead. As $u$ and $v$ have the same neighborhood, the independent set condition is not violated and neither is the requirement of moving tokens only along edges. That is, if there is no token at $v$ at that same step. However, a token on $v$, blocking all of $N(u)$, would prohibit the placement of a token on $u$.
\end{proof}

\begin{lemma}
	\label{lem:kernel-exsistence}
	\isdts admits a kernal on DAGs of depth 3 when parameterized by~$k$.
\end{lemma}
\begin{proof}
	We first use \cref{lem:dag-depth-3-regular} to obtain an instance, where the first layer is equal to $S$ and the third layer to $D$. Then we apply \cref{rule:not-the-same-neighbourhoods}. Note, that now every vertex on the second layer has a distinct neighborhood which is a subset of $S\cup D$. 
	As there are at most $2^{|S\cup D|}$ unique subsets, the total number of vertices is bound by $|S|+|D|+2^{|S\cup D|} = 2k+4^k$.
\end{proof}
 
Together \cref{lem:kernel-exsistence} and \cref{lem:poly-solveable} complete the proof of \cref{theo:tractability}.
	
	\section{NP completeness for DAGs of depth 3}\label{sec:np-hard}
	
In this section we turn to hardness results, improving the lower bounds of~\cite{ito2022}.
While they prove that \isdts is \NP-hard on DAGs, their reduction yields DAGs of unbounded depth.

\npcompletedepthtwo*

\begin{figure}[H]
	\center
	\includegraphics[]{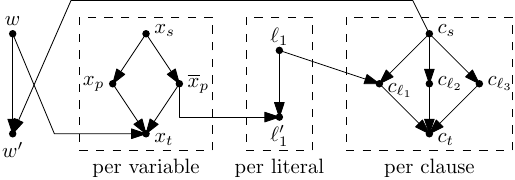}
	\caption{Schematic of the graph resulting from a 3-SAT formula, where the clause $c$ contains the positive literal $x$.}\label{pic:NP}
\end{figure}

The proof of \cref{theo:np-complete-depth-2} is obtained by reducing from 3-SAT. For a 3-SAT sentence $\varphi$ we construct the following instance $(G_\varphi, S_\varphi, D_\varphi)$:
\begin{enumerate}
	\item Add a timing gadget, meaning two vertices $w \in S_\varphi$ and $w'\in D_\varphi$ as well as an edge $ww'$. 
	\item For each variable $x$ create four vertices $x_s \in S_\varphi$, $x_t \in D_\varphi$, $x_p$ and $\overline{x}_p$ as well as edges $x_sx_p$, $x_s\overline{x}_p$, $x_px_t$ and $\overline{x}_px_t$. Further, connect this subgraph to the timing gadget with the edge $wx_t$.
	\item For each clause $c$ with the literals $\ell_1,\ell_2,\ell_3$ add five vertices $c_s \in S_\varphi$, $c_t \in D_\varphi$ and $c_{\ell_1},c_{\ell_2},c_{\ell_3}$. Further, add edges form $c_s$ to $c_{\ell_1},c_{\ell_2} $ and $c_{\ell_3}$ as well as from $c_{\ell_1},c_{\ell_2} $ and $c_{\ell_3}$ to $c_t$. Additionally, connect this subgraph to the timing gadget with the edge $c_sw'$.
	
	\item For every literal $\ell$ add two vertices $\ell \in S_\varphi$ and $\ell' \in D_\varphi$ and edge $(\ell,\ell')$. For all $\overline x = \ell$ (note that $\overline{\overline{x}}=x$) and $c$ containing $x$ add edges $(\ell, c_\ell)$ and $x_p\ell'$ respectively.
\end{enumerate}
This construction is represented in \cref{pic:NP}

\begin{lemma}
	If $\varphi$ is satisfiable then $(G_\varphi, S_\varphi, D_\varphi)$ is a positive instance for \isdts.
\end{lemma}
\begin{proof}
	Let $\sigma$ be a satisfying assignment of $\varphi$.
	We define the following reconfiguration sequence. In no particular order move a token from $x_s$ to 
	$x_p$ if $\sigma(x)=$ True, to $\overline{x}_p$ otherwise. 
	
	For all literals where it is possible without violating the independent set condition, slide along $(\ell,\ell')$. I.e.~this move is possible for every literal of the form $x$ (resp. $\overline{x}$) where $\sigma$ sets $x$ to True (resp. to False).
	
	As each clause $c$ has at least one literal $\ell$ that evaluates to true. Move the token from $c_s$ to $c_t$ over $c_{\ell}$.
	This move is possible as the token in the ``literal gadget'' moved from $\ell$ to~$\ell'$.
	Now one is able to move the token from $w$ to $w'$.
	Next, move the tokens on all $x_p$ or $\overline{x}_p$ to~$x_t$ and finally from all remaining $\ell$ to $\ell'$.
\end{proof}

\begin{lemma}
	If $(G_\varphi, S_\varphi, D_\varphi)$ is a positive-instance of \isdts then $\varphi$ is satisfiable.
\end{lemma}
\begin{proof}
	We first show, that the flow of token is tightly controlled.
	\begin{claim}
		\label{lem:token-flow-in-a-controlled-way}
		The token coming from $w$ will only slide along the edge $ww'$. Further, for all variables $x$, clauses $c$ and literals $\ell$ the tokens from $x_s$, $c_s$ and  $\ell$ will end up in $x_t$, $c_t$ and $\ell'$ respectively.
	\end{claim}
	\begin{claimproof}
		By reachability in $G_\varphi$ the only possibility where this is not the case would be if some token coming from $x_s$ ends up in $\ell'$ (for $x = \ell$). Implying the token from $\ell$ ends up in $c_t$; implying that the token from $c_s$ ends up in $w'$; implying that the token from $w$ slides to $x_t$. However, as $w$ blocks $w'$, $c_s$ blocks $c_\ell$, $\ell$ blocks $\ell'$ and $x_t$ locks in $x_s$, so none of these moves is doable.
	\end{claimproof}
	
	Let $(\alpha_i)_{i\in I}$ be a reconfiguration sequence certifying that $(G_\varphi, S_\varphi, D_\varphi)$ is a positive-instance. The previous claim implies
	that the token in $w$ ends up in $w'$, so let $\alpha_i$ be the first configuration where a token occupies $w'$. We define the following set $X' = \{\ell' \in \alpha_i(T) \mid \ell \textrm{ is literal}\}$.
	
	\begin{claim}
		The set $X'$ contains one literal from each clause. Further, if $\ell$ is in $X'$, then $\overline{\ell}$ is not.
	\end{claim}
	\begin{claimproof}
		By \cref{lem:token-flow-in-a-controlled-way}, for all clauses $c$ the token coming from $c_s$ is either at some $c_\ell \in \{c_{\ell_1}$, $c_{\ell_2}, c_{\ell_3}\}$ or is already on $c_t$ and has previously been in one of the $c_\ell$. Thus, no token can be at its neighbor $\ell$ (on the literal gadget). By \cref{lem:token-flow-in-a-controlled-way} there is now a token on~$\ell'$.
		
		For the second claim, assume towards a contradiction, that in $\alpha_i$ tokens are placed at both $\ell$ and $\overline{\ell}$ for some literal $\ell$. This implies that neither for $\alpha_i$ nor for any succeeding configuration a token is at $x_p$ or $\overline{x}_p$ where $x$ is the variable of the literal $\ell$. As no token is on~$x_t$ there is one at $x_s$ and by \cref{lem:token-flow-in-a-controlled-way} it cannot reach a valid end position.
	\end{claimproof}
	
	We find the following assignment $\sigma()$. Set $\sigma(x) = $ True, if $x \in X'$ and otherwise $\sigma(x) = $ False. Observe, that all literals in $X'$ evaluate to true. As $X'$ contains at least one literal per clause all clauses evaluate to true.
\end{proof}

This concludes the proof of \cref{theo:np-complete-depth-2}.
	
    \section{W[1]-hardness for DAGs of Depth 4}\label{sec:w1-hard}
    
In this section, we prove the second lower bound of this paper. Here, $k$ is the number of tokens.

\woneharddepthfour*

We show this result by providing a \FPT-reduction from Independent Set. That is given an instance for independent set (with parameter $k$), we build an instance of \isdts with parameter $f(k)$, such that the input instance is a positive instance if and only if the resulting one is. \cref{sec:construction} describes the construction while \cref{lem:corrrect-w1} in \cref{sec:correctness} proves the correctness of the reduction.

\subsection{Construction}\label{sec:construction}
\begin{figure}[h]
	\centering
	\tikzset{subgraphl/.style={rectangle, draw=black, minimum width=3cm}}
	\tikzset{subgraph/.style={rectangle, draw=black, minimum width=2.5cm}}
	\tikzset{node/.style={circle, draw=black, minimum width=0.5cm, fill=white}}
	
	\tikzset{multiconnect/.style={-{Latex[fill=white]},double, double distance = 0.05cm, thick}}
	\includegraphics[width=\linewidth]{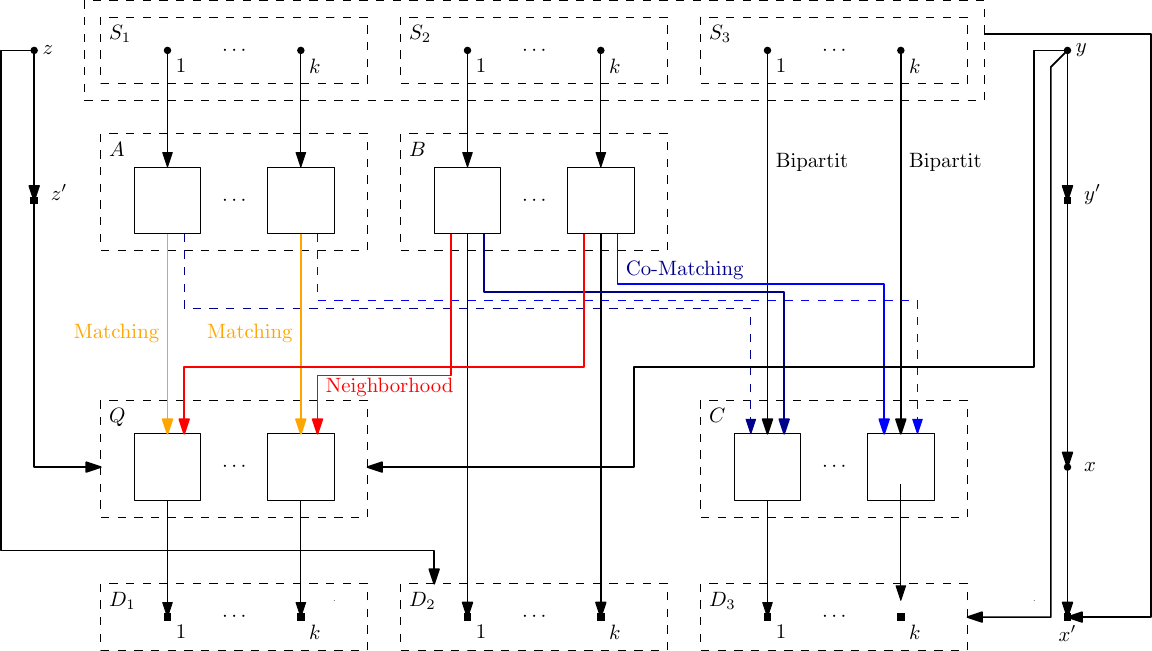}
	\caption{Conceptual view on the graph resulting from the reduction. Arrows represent directed edges between vertices in the two blocs they connected. Unless labeled otherwise the connected is complete. Dashed arrows have no different meaning and are only stylized for clarity.}
	\label{fig:conceptual-view}
\end{figure}

\newcommand{\component}[1]{\ensuremath{\textrm{\underline{#1}}}}
The reduction takes as input an undirected graph $G$ of $n$ vertices and an integer $k$ and outputs the graph as described in the following. This graph is also depicted in \cref{fig:conceptual-view}.

\medskip\noindent{\bf Vertices.}
In total the DAG consists of 10 main blocs and three helping gadgets. Let us commence with the main blocs. Six of these consist of $k$ isolated vertices. Half of the blocs: $S_1$, $S_2$ and $S_3$ contains only starting positions: $s_{1,1}$ to $s_{3,k}$. The other half: $D_1$, $D_2$, and $D_3$ only contain target (or destination) vertices: $d_{1,1}$ to $d_{3,k}$.

The remaining four components are each made up of $k$ copies of the vertex set of $G$, i.e.~$kn$ isolated vertices. The components are named $A$, $B$, $C$, $Q$, and, given $i\le k$, $v\in V(G)$, we call $a_{i,v}$ (resp. $b_{i,v}$, $c_{i,v}$, $q_{i,v}$) the copy of $v$ in the $i$th copy $A_i$ of $V(G)$ in $A$ (resp.~$B$,~$C$,~$Q$).

Lastly, we have six specific vertices, controlling the flow of tokens in the graph: $x$, $x'$, $y$, $y'$, $z$, $z'$, where $x,y,z$ are starting positions, and $x',y',z'$ destinations.

\medskip\noindent{\bf Edges.}
The edges of our construction are represented in \cref{fig:conceptual-view} and are as follows.
\begin{enumerate}
	\item \underline{Full connections:} For every $i\le k$ and $v$ in $ V(G)$, connect $s_{1,i}$ (resp. $s_{2,i}$, $s_{3,i}$) to $a_{i,v}$ (resp. $b_{i,v}$, $c_{i,v}$).
	Further, connect $q_{i,v}$ (resp. $b_{i,v}$, $c_{i,v}$) to $d_{1,i}$ (resp. $d_{2,i}$, $d_{3,i}$).
	\item \underline{Matching connections}: For every $i\le k$, and $v$ in $V(G)$, connect $a_{i,v}$ to $q_{i,v}$. I.e. make a matching between $A_i$ and $Q_i$.

	\item \underline{Co-matching connections:} For every $i\le k$, and $u,v$ in $V(G)$ if and only if $v\neq u$, connect  $a_{i,v}$ to $c_{i,u}$ and connect $b_{i,v}$ to $c_{i,u}$. I.e. Make a co-matching between $A_i$ and $C_i$ and between $B_i$ and $C_i$.
	
	\item \underline{Neighborhood connections:} For every $i,j\le k$, and $u,v$ in $V(G)$ if and only if $i\neq j$ and $u\in N^G(v)$, connect $b_{i,v}$ to $q_{j,u}$.
	
	\item \underline{Gadget connections:} Connect $x$ and each vertex in $S_1$, $S_2$, and $S_3$ to $x'$. Connect $y'$ to $x$. Further, connect $y$ to $y'$, and to each vertex in $Q$ and $D_3$. Lastly, connect $z$ to $z'$ and to each vertex in $D_2$, and connect $z'$ to each vertex in $Q$.
\end{enumerate}

We define the start positions as $S:= S_1\cup S_2\cup S_3 \cup \{x,y,z\}$ and destination positions as~$D:=D_1\cup D_2\cup D_3 \cup \{x',y',z'\}$.

We call our resulting DAG $H$. Notice that $H$ has $4kn+6k+6$ many vertices, and $3k+3$ start / destination vertices. Furthermore, the DAG has depth 4 has illustrated in \cref{fig:conceptual-view}.

\subsection{Correctness}\label{sec:correctness}
This section is devoted to the proof of the correctness of the reduction, as stated in the following lemma.

\begin{lemma}\label{lem:corrrect-w1}
	If the input $G,k$ is a positive instance for the independent set problem if and only if $(H,S,D)$ is a positive \isdts instance.
\end{lemma}

We now prove the first implication. Assume that $X=\{v_1,\ldots,v_k\}\subseteq V(G)$ is an independent set of $G$ of size $k$. We move the tokens in $H$ as follows.
\begin{enumerate}
	\item For every $i\le k$, move the token from $s_{3,i}$ to $c_{i,v_i}$. Similarly, move tokens from $s_{1,i}$ and $s_{2,i}$ to $a_{i,v_i}$, and $b_{i,v_i}$ respectively.
	\item Move the token from $x$ to $x'$ and the token from $y$ to $y'$.
	\item For every $i\le k$, move the token from $a_{i,v_i}$ to $q_{i,v_i}$. And then from $q_{i,v_i}$ to $d_{1,i}$.
	\item Move the token from $z$ to $z'$.
	\item For every $i\le k$, move the token from $b_{i,v_i}$ to $d_{2,i}$ and the tokens from $c_{i,v_i}$ to $d_{3,i}$.
\end{enumerate}

The first set of moves is possible because a token on $c_{i,v_i}$ blocks in $A_i$ every $a_{i,u}$ with $u\neq v_i$, but $a_{i,v_i}$ is free. The same holds for $b_{i,v_i}$.
Second, the clocks $x,y$ can move to $x',y'$ because $S_1$, $S_2$, and $S_3$, are free of token, freeing $D_3$.
Third, $X$ is an independent set in $G$, and therefore $v_i\in X$ is not adjacent to any $v_j \in X$ for $j\neq i$. Therefore, $q_{i,v_i}$ is not adjacent to any $b_{j,v_j}$ for $j\neq i$. Hence, the tokens can move from $A$ to $Q$ and then to $D_1$.
Forth, as~$Q$ is empty of token, $z$ can move to $z'$, freeing $D_2$.
Last, the tokens in $B$ and $C$ can freely move to $D_2$ and $D_3$ respectively.

This shows that $(H,S,D)$ is a positive \isdts instance, and conclude the first part of the proof of \cref{lem:corrrect-w1}.
The second implication in the statement of \cref{lem:corrrect-w1} is harder to prove. We show that any reconfiguration sequence must follow some structure, from which we derive the existence of an independent set in $G$. In what follows, we assume that $(H,S,D)$ is a positive \isdts instance. 

\begin{claim}
	For any reconfiguration sequence, the tokens in $x$, $y$, and $z$ end up in $x'$, $y'$, and $z'$ respectively. Furthermore, they move in this order. 
\end{claim}
\begin{claimproof}
	While other tokens can reach $x'$, the token on $x$ can only reach $x'$. Also, the token in~$y$ (resp. $z$) is the only one that can reach $y'$ (resp. $z'$).

	Additionally, notice that both $y$ and $z'$ block all of $Q$. As the token reaching $D_1$ need to go through $Q$, if $z$ moves to $z'$ before $y$ moves to $y'$ no token can ever reach $Q$. So $y$ must reach $y'$ before $z$ reaches $z'$. And since $x$ blocks $y'$, $x$ must slide to $x'$ first.
\end{claimproof}

\begin{claim}
	\label{claim:tightly_controlled_paths}
	For any reconfiguration sequence, no token slides along the edge $uv$, if $uv$ is in $A\times C$, $B \times C$ or $B\times Q$.
\end{claim}
\begin{claimproof}
	As the tokes coming from the clocks stay within them, we may argue as if these vertices and their respective tokens were not present.
	
	The vertices in $D_2$ can only be reached from $S_2$, thus no token may leave a path from some vertex in $S_2$ to some vertex in $D_2$ which prohibits the use of edges between $B$ and $C$ or $B$ and $Q$. Consequently and similarly, the only tokens that can end up in $D_1$ are these that start in $S_1$, prohibiting the use of the edges between $A$ and $C$. 
\end{claimproof}

\begin{claim}
	There exists a set $X=\{v_1,\ldots v_k\}\subseteq V(G)$ such that when $y$ moves to $y'$, there are tokens on vertices $a_{i,v_i}$, $b_{i,v_i}$, and $c_{i,v_i}$ for every $i\le k$.
\end{claim}
\begin{claimproof}
	When $y$ moves to $y'$, $x$ already moved to $x'$, so $S_1$, $S_2$, and $S_3$ are empty of tokens. Furthermore, as $y$ and $z$ did not move yet, $D_2$, $D_3$, $Q$ and therefore $D_1$ are empty of token. So the tokens in $S_3$ moved to $C_3$. We then define $v_i$ as the vertex of $V(G)$ such that $s_{3,i}$ moved to $c_{i,v_i}$. Finally, note that $c_{i,v_i}$ blocks in $A_i$ every $a_{i,u}$ with $u\neq v_i$. So the token from $s_{1,i}$ must now be in $a_{i,v_i}$, as these tokens may not move to other blocs by \cref{claim:tightly_controlled_paths}.  Similarly, the token from $s_{2,i}$ must now be in $b_{i,v_i}$.
\end{claimproof}

\begin{claim}
	$X$ is an independent set of size $k$ in $G$. 
\end{claim}
\begin{claimproof}
	When $z$ goes to $z'$, all of $Q$ is blocked, and remains so for the rest of the reconfiguration. So the vertices in $D_1$ are already occupied with tokens. These tokens must be from $A$ by \cref{claim:tightly_controlled_paths}.
	
	So before $z$ goes to $z'$ each token in $a_{i,v_i}$ must go to $q_{i,v_i}$ (there is only one outgoing edge from $a_{i,v_i}$), and then to $d_{1,i}$.
	During these moves, the tokens in $B$ remains there since $D_2$ is still blocked by $z$.
	If $a_{i,v_i}$ can freely move to $q_{i,v_i}$ it implies that no token on $b_{j,v_j}$ is adjacent to $q_{i,v_i}$ which implies that no $v_j$ (with $j\neq i$) is adjacent to (or equals) $v_i$ in $G$. As this holds for every $i\le k$, we have that $X$ is an independent set in $G$.
\end{claimproof}

In conclusion, the existence of a reconfiguration sequence implies the existence of an independent set in $G$. This concludes the proof of \cref{lem:corrrect-w1}, and therefore of \cref{theo:w1-hard-depth-4}.

	\section{FPT when parameterized by Treewidth and $k$}\label{sec:tw}
	Now that we have covered the impact of the depth of a DAG on the tractability of \isdts, we look at other subfamily of DAGs that could enable \FPT\xspace algorithms. Recently, Ito et al.\cite[Theorem 3]{ito2022} proved that \isdts is polynomial time solvable on DAGs whose underlying graph is a tree.
We extend this results by looking at graph of bounded treewidth.

\fpttwk*

We first prove several results on galactic reconfiguration and the impact of collapses that we defined in \cref{sec:prelim}. Then, we explain how to perform a dynamic algorithm on a tree decomposition.

\subsection{Reconfiguration and Collapses}

The first lemma shows that a reconfiguration defined on a graph only has one possible collapse on a collapse of the graph. Meaning that it does not matter in which sequence of collapse the new graph is obtained. 
\begin{lemma}
	\label{lem:transitivity}
	Let $G_1$, $G_2$, $G_2'$, and $G_3$ galactic graphs such that both $G_2$ and $G_2'$ are obtained from $G_1$ through collapses, and $G_3$ is obtained from both $G_2$ and $G_2'$ through collapses.
	
	Let $\alpha,\beta,\beta',\gamma$ be reconfiguration sequences over $G_1,G_2,G'_2,G_3$ respectively. If $\alpha$ collapses to $\beta$, $\alpha$ collapses to $\beta'$ and $\beta$ collapses to $\gamma$, then $\beta'$ collapses to $\gamma$.
\end{lemma}

\begin{proof}
	 Let $\gamma'$ be the sequence $\beta'$ collapses to in $G_3$. We show that $\gamma = \gamma'$.
	 
	Let $H_1,H_2,\dots,H_{|I|}$ be the sequence of graphs in the chain of collapses from $G_1$ over $G_2$ to $G_3$. For each $i\in I\setminus\{|I|\}$ let $f_i$ denote the function that maps vertices (planets and black holes) from $H_i$ to vertices $H_{i+1}$ they correspond to. That is, to itself if they where not collapsed and to the created black hole otherwise.
	
	We have to account for the deletion of consecutive equal configurations, when collapsing reconfiguration sequences. Therefore, let $g_i$ be the function that maps indices of the reconfiguration sequence that $\alpha$ collapses to in $H_i$ to the corresponding indices in the reconfiguration sequence that $\alpha$ collapses to in $H_{i+1}$.

	Let $f = f_1 \circ f_2 \circ \dots \circ f_{|I|-1}$ and define $g$ analog. Further, define $f'$ and $g'$ analog for the chain of collapses in from $G_1$ to $G_2'$ and finally $G_3$.
	By definition $\gamma_{g(j)}(t) = f(\alpha_j(t))$ for each configuration $\alpha_j$ and each token $t$. Analog, the same holds for $\gamma'$ with $f'$ and $g'$.

	As the resulting graph for both chains of collapses is the same, $f = f'$. Further, as equality classes of configurations only merge when collapsing, we may assume that consecutive equal configurations get resolved only at the end of a chain of collapses. Thus as $f=f'$ also $g=g'$. Thus in conclusion, for all token $t$ and steps $g(i)$: $\gamma'_{g(i)}(t) = f'(\alpha_i(t)) = f(\alpha_i(t)) = \gamma_{g(i)}(t)$.
\end{proof}

Observe that these lemma can be iterated to obtain the same statement for arbitrary long chains of collapses, essentially proves that the collapsing is transitive.

The next lemma shows that on DAGs, reconfiguration are restricted.
\begin{lemma}
	\label{lem:only-unique-steps}
	Let $\alpha$ be a reconfiguration sequence over a DAG $G$ without back holes. Then for every $i<j<\ell$ and every token $t$ we have that $\alpha_i(t) = \alpha_\ell(t)$ implies $\alpha_i(t)=\alpha_j(t)$.

	Furthermore, if $\beta$ is a collapse of $\alpha$, then for all $i<j<\ell$ and token $t$ such that $\beta_i(t) = \beta_\ell(t)$ and $\beta_i(t)\neq\beta_j(t)$, we have that $\beta_i(t)$ is a black hole.
\end{lemma}
\begin{proof}
	The first assertion is trivial. A token returning to a vertex would imply a round walk and thus a cycle.
	Assume that $\beta_i(t) = \beta_\ell \in A(G)$ there exists corresponding indices $i'<j'<\ell'$ with $\alpha_{i'}(t) = \alpha_{\ell'}(t) \neq \alpha_{j'}(t)$ -- contradicting the first statement.	
\end{proof}

This in turn bound the number of possible reconfiguration sequences that can be on a DAG or its collapses.
\begin{lemma}
	\label{lem:upper-limit-for-sequence-lengths}
	For a DAG without black holes $G$ and a graph $G'$ obtained by collapses from $G$ where additionally $G'$ does not contain neighboring black holes; the number and length of reconfiguration sequences over $G'$ that are collapses of reconfiguration sequences over $G'$ is bounded by $|V(G')|^{2k^2|V(G')|}$.
\end{lemma}
\begin{proof}
	No token can return to a planet after leaving it, due to \cref{lem:only-unique-steps}. A token might be at a black hole multiple times. However, as there are no repeating consecutive configurations and by assumption no neighboring black holes, a change for a planet has to happen in between consecutive configurations. 
	This bounds the length of every sequence by $2k|V(G')|$.

	As there are only $|V(G')|^k$ possible configurations the total number of possible reconfiguration sequences is at most $(|V(G')|^k)^{2k|V(G')|} = |V(G')|^{2k^2|V(G')|}$.
\end{proof}

Our last lemma of this section is a composition lemma that explains how to glue together two reconfiguration sequences defined on distinct collapses of a given graph.
\begin{lemma}[Composition lemma]
	\label{lem:gluing}
	Let $G$ be a graph and $U,V$ disjoint sets of vertices such that $U$ has no neighbors in $V$ and vise versa. Let $(\alpha_i)_{i\in I}$ and $(\beta_j)_{j\in J}$ be reconfiguration sequences, for $G\odot U$ and $G\odot V$ respectively, that collapse to the same sequence in $G\odot U \odot V$. Then there exists a reconfiguration sequence in $G$ that collapses to $(\alpha_i)_{i\in I}$ and $(\beta_j)_{j\in J}$ over $G\odot U$ and $G\odot V$ respectively.
\end{lemma}
\begin{proof}
	Let $G,U,\alpha = (\alpha_i)_{i\in I}$ and $\beta = (\beta_j)_{j\in J}$ defined as above. Let $b_U$ and $b_V$ the black holes in $G\odot U \odot V$ corresponding to the sets $U$ and $V$ respectively and let $ \gamma = (\gamma_\ell)_{\ell \in L}$ be the shared collapsed reconfiguration sequence in $G\odot U \odot V$. 
	
	We build a configuration sequence $(\delta_d)_{d\in D}$ over $G$ by concatenating sequences $\delta^\ell$ (for each $\ell\in L$).  

	Broadly speaking, fix $\ell\in L$ and the configuration $\gamma_\ell$, and look at the configurations in $(\alpha_i)_{i\in I}$ and $(\beta_j)_{j\in J}$ that correspond to it. All these configurations in $\alpha$ are equivalent when restricted to $G\odot U \odot V$, therefore tokens only move in $V$ (and are all mapped to $b_V$ in $\gamma_\ell$). Similarly, the configuration in $\beta$ corresponding to this $\gamma_\ell$ only differ in $U$. So in $\delta$ we can perform all these move independently before moving to the configuration $\gamma_{\ell+1}$.
	
	Formally let $I_\ell\subseteq I$ the set of index of $(\alpha_i)_{i\in I}$ corresponding to $\gamma_\ell$, and define $J_\ell$ similarly.
	Let $i_\ell$ be the minimum index of $I_\ell$ minus one
	and let $j_\ell$ be the minimum index of~$J_\ell$. Let $\delta^\ell$ be the sequence of length $|I_\ell|+|J_\ell| -1$ such that for every integer~$d$ and token~$t$, we have:
	\begin{equation*}
		\delta^\ell_d (t) := 
		\begin{cases}
			\begin{array}{lll}
				\alpha_{i_\ell + d}(t) &\text{ if } \alpha_{i_\ell + d}(t)\neq b_U  &\text{ and if } d\le |I_\ell|\\
				
				\beta_{j_\ell}(t) & \text{ if } \alpha_{i_\ell + d}(t) = b_U   &\text{ and if } d\le |I_\ell| \\
				
				\beta_{j_\ell + d-|I_\ell|}(t) & \text{ if } \beta_{j_\ell + d-|I_\ell|}(t)\neq b_V  &\text{ and if } d > |I_\ell|\\
				
				\alpha_{i_\ell+|I_\ell|}(t) & \text{ if } \beta_{j_\ell + d-|I_\ell|}(t) = b_V  &\text{ and if } d> |I_\ell|
			\end{array}
		\end{cases}
	\end{equation*}
	
	Note that the definition is consistent as $\alpha_i(t) = \beta_j(t) = \gamma_\ell(t)$ for every $i\in I_\ell$ and every $j\in J_\ell$ as soon as $\gamma_\ell(t)$ is not one of the black hole $b_U$ or $b_V$.
	Finally, define $\delta$ as the concatenation of every $\delta^\ell$ for $\ell\in L$.

	To see that $\delta$ is indeed reconfiguration sequence, note that each time only one token moves (according to the move defined in $\alpha$ or $\beta$). Furthermore, since $U$ and $V$ are disjoint and non-adjacent the set of tokens has to be an independent set (when only considering planets). Otherwise, this inconsistency already holds in $\alpha$ or $\beta$. 

	By construction, $\delta$ and $\alpha$ (resp.~$\beta$) agree on the positioning of tokens except for $b_U$ (resp.~$b_V$). And therefore $\delta$ collapses to $\alpha$ (resp.~$\beta$) on $G\odot U$ (resp.~$G\odot V$). 
\end{proof}

\subsection{Dynamic Programming}\label{sec:dyn-prog}

In a nutshell, our algorithm will try every possible reconfiguration sequence for each bag. Starting from the leaf, the information of what are the possible reconfiguration sequences is computed by a bottom up manner. At every node $x$ the algorithm tries to glue the possible reconfiguration sequences from the left and from the right children, checking that they agree on the overlapping part. Thanks to \cref{lem:upper-limit-for-sequence-lengths} we only have to consider small sequences at the bag level, and the gluing can be done thanks to \cref{lem:gluing}. We now turn to formal definitions.
We consider a DAG $G$, and it's tree-decomposition as defined in \cref{sec:prelim}.

\begin{figure}[H]
	\begin{center}
		\includegraphics[width=0.9\linewidth]{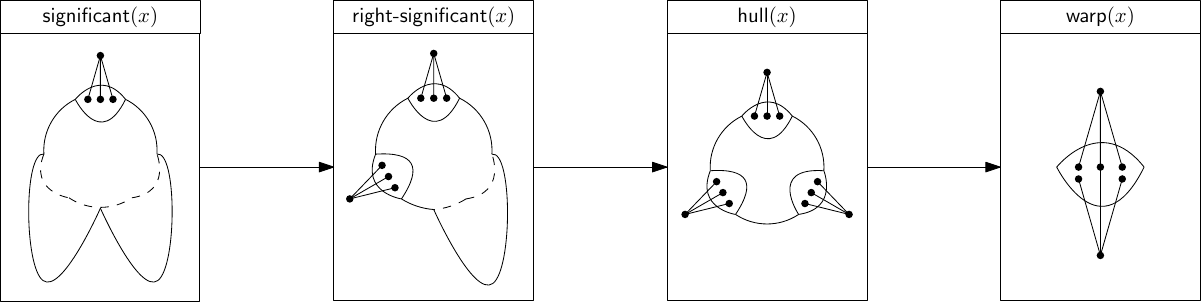}
	\end{center}
	\caption{Collapsing different graphs into each other.}\label{pic:ex-sig}
\end{figure}

\begin{figure}[H]
	\includegraphics[width=\linewidth]{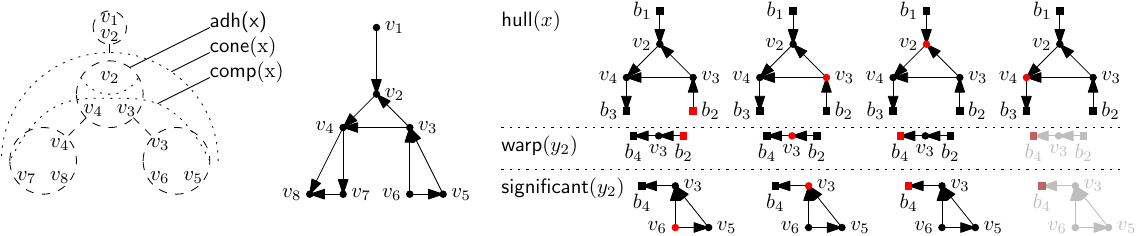}
	\caption{A graph, its tree-decomposition (left) and reconfiguration sequences with one token (red) over collapsed graphs (right), where $y_2$ is the right child of $x$. The reconfiguration over $\hull(x)$ and $\significant(y_2)$ can be glued together using \cref{lem:gluing} as they collapse to the same sequence over $\warp(y_2)$. The reconfiguration sequences are presented with repeating configuration for clarity.}\label{pic:ex-explicit}
\end{figure}

Let $x$ be a node of the tree-decomposition and $y_1, y_2$ its children. We define the {\em significant} of $x$ noted $\significant(x)$ as $G \odot (V\setminus\cone(x))$. Intuitively, everything that is ``above'' $x$ in the tree decomposition has been collapsed into one black hole.
The {\em right-significant}, noted $\rightsignificant(x)$ as $\significant(x) \odot \cmp(y_1)$ and $\hull(x)$ as the graph $\rightsignificant(x) \odot \cmp(y_2)$. I.e.~where one (or both) subtrees of $x$ have collapsed into a black hole. Finally, we define $\warp(x)$ as $\hull(x)\odot\mrg(x)\cup\{b_1, b_2\}$ where $b_1$ (respectively $b_2$) is the black hole obtained from collapsing $\cmp(y_1)$ (resp $\cmp(y_2)$).
Note that $\warp(x) = \significant(x) \odot (\cmp(x))$. I.e.~it is only composed of element of the adhesion of $x$ with two black holes representing everything that is ``above'', and ``bellow'' in the tree decomposition.
Observe that $\hull(x)$ can be collapsed to obtain all $\warp(x)$, $\warp(y_1)$ and $\warp(y_2)$ in a single step.
All these notions are represented in \cref{pic:ex-sig}, and an example of reconfiguration is presented in~\cref{pic:ex-explicit}.

\begin{definition}
	Let $G$ be a DAG, together with a tree decomposition, let $x$ be a node of the tree decomposition. The {\em profile} of $x$ is the set of reconfiguration sequences $\alpha$ over $\warp(x)$ such that there is a reconfiguration $\beta$ over $\significant(x)$ that collapses to $\alpha$ over $\warp(x)$.
\end{definition}
Thanks to \cref{lem:upper-limit-for-sequence-lengths}, the profile of each node is bounded by $\tw^{2k\cdot\tw}$, where $\tw$ is the width of the given tree decomposition. Ne now show that we can compute the profile of each node in a bottom up manner.

\begin{proposition}\label{prop:dp-algo}
	Given a tree decomposition of a DAG $G$ of width \tw, a node $x$ of the decomposition, and the profile of its children $y_1$ and $y_2$, we can compute the profile of $x$ in time $f(k,\tw)$, for some computable function $f$.
\end{proposition}
\begin{proof} 
For the leaf of the tree-decomposition, it holds trivially, as we can compute all possible sequences in $\significant(x)$ as it is of size at most $\tw$. Thus, we may assume that $x$ is not a leaf. For every possible reconfiguration sequence $\alpha$ over~$\warp(x)$ the algorithm adds it to the profile of $x$ if and only if the following holds.

There are reconfiguration sequences $\beta_1$, $\beta_2$, and $\gamma$ over $\warp(y_1)$, $\warp(y_2)$, and $\hull(x)$ respectively, such that :
\begin{enumerate}
	\item $\beta_1$ (resp. $\beta_2$) belongs to the profile of $y_1$ (resp. $y_2$), and
	\item $\gamma$ collapses to $\alpha$ (resp. $\beta_1$, $\beta_2$ ) over $\warp(x)$ (resp. $\warp(y_1)$, $\warp(y_2)$).
\end{enumerate}

\begin{claim}
	This checks can be performed in time $f(k,\tw)$.
\end{claim}
\begin{claimproof}
	By \cref{lem:upper-limit-for-sequence-lengths}, there can be only few possible sequences to consider for $\gamma$. We can then compute its collapses in $\warp(y_1)$ and $\warp(y_2)$, and check that the results belongs to the profiles of $y_1$ and $y_2$ that have already been computed. All of this can therefore be computed in time $f(k,\tw)$.
\end{claimproof}
 
\begin{claim}
	If there exists a reconfiguration sequence $\delta$ over $\significant(x)$ that collapses to~$\alpha$ in $\warp(x)$ then the algorithm adds $\alpha$ to the profile of $x$.
\end{claim}
\begin{claimproof}
	We first collapse $\delta$ to a reconfiguration $\gamma$ in the graph $\hull(x)$.
	By \cref{lem:transitivity} we obtain that $\gamma$ also collapses to $\alpha$ over $\warp(x)$.
	
	Now, we collapse $\delta$ to a sequence $\beta'_1$ over $\significant(y_1)$, and both collapse to a sequence~$\beta_1$ over $\warp(y_1)$. Since $\beta'_1$ is in the significant of $y_1$, $\beta_1$ is in the profile of $y_1$. We define~$\beta'_2$ and~$\beta_2$ similarly for $y_2$.
	
	Note that this triple ($\gamma$, $\beta_1$, and $\beta_2$) must be considered by the algorithm. Finally, by \cref{lem:transitivity}, $\gamma$ also collapses to $\beta_1$ and $\beta_2$. So this triple of reconfiguration sequences passes the checks, and the algorithm selects $\alpha$ to the profile of $x$.
\end{claimproof}

\begin{claim}
	If the algorithm adds $\alpha$ to the profile of $x$, then there exists a reconfiguration sequence $\delta$ in $\significant(x)$ that collapses to $\alpha$ in $\warp(x)$
\end{claim}
\begin{claimproof}
	Let $\gamma$ be the guessed sequence in $\hull(x)$ and  $\beta_1,\beta_2$ be the sequences that $\gamma$ collapses to in $\warp(y_1)$, $\warp(y_2)$ respectively, where $y_1$ and $y_2$ are the left and right children of~$x$.
	As $\beta_1$ and $\beta_2$ are in the profile of $y_1$ and $y_2$ respectively,
	by definition of profile, there exist~$\beta'_1$ and~$\beta'_2$ in $\significant(y_1)$ and $\significant(y_2)$ that collapse to $\beta_1$ and $\beta_2$ in $\warp(y_1)$ and $\warp(y_2)$ respectively.
	
	Now we use the Composition \cref{lem:gluing} on $\gamma$ and $\beta'_2$ to first get a reconfiguration sequence~$\gamma'$ in $\rightsignificant(x)$.
	This is possible because they both collapse to $\beta_2$ according to the check of the algorithms, and because by construction the vertices in $\cmp(y_2)$ are only adjacent to the vertices of $\adh(y_2)$ (which are in $\warp(y_2)$) and not to the other vertices of $\rightsignificant(x)$.

	As $\gamma'$ collapses to $\gamma$, and the algorithm checked that $\gamma$ collapses to $\beta_1$, by \cref{lem:transitivity}, $\gamma'$ also collapses to $\beta_1$.
	We can finally use again our Composition \cref{lem:gluing}, because both $\beta'_1$ and $\gamma'$ collapse to $\beta_1$ over $\warp(y_1)$.
	And as before, the vertices of $\significant(y_1)$ that are not in $\warp(y_1)$ (i.e.~the vertices of $\cmp(y_1)$) are only adjacent to the vertices of $\adh(y_1)$, which are in $\warp(y_1)$, and not to the other vertices of $\significant(x)$. This yields a reconfiguration sequence $\delta$ over $\significant(x)$. With several uses of \cref{lem:transitivity}, we obtain that $\delta$ which collapses to $\gamma'$ also collapses to $\gamma$, and therefore also collapses to $\alpha$.
	
	In conclusion, this sequence $\alpha$ is indeed in the profile of $x$.
\end{claimproof}

This therefore concludes the proof of \cref{prop:dp-algo}.
\end{proof}

With all of that in mind, we can also conclude the proof of \cref{theo:fpt-tw-k}.
Thanks to \cref{prop:dp-algo}, we compute the profile of every node in time linear in the number of nodes in the tree decomposition (with constant factors depending on $k$ and $\tw$).
As the significant of the root is the entire graph, the non-emptiness of the profile of the root certifies the existence of a reconfiguration sequence in the entire graph.
Such a reconfiguration sequence can be computed in a top-down traversal of the tree decomposition,
selecting at each level a sequence from the profile that is compatible with the one selected by the parent.

	\section{Back to undirected graphs}\label{sec:undirected}
	
One of the biggest open problem in token sliding reconfiguration
is whether \ists parameterized by $k$ is \FPT\xspace on graphs of bounded treewidth (see \cite[Question 5]{Bousquet2024}). Our study of directed graphs and \cref{theo:fpt-tw-k} can be seen as a step towards solving this open problem. In fact the techniques that we develop in \cref{sec:tw} have consequences for undirected graphs. 

For example, note that the only place in the proof of \cref{theo:fpt-tw-k} where we use that the given graph is directed and is a DAG is in \cref{lem:only-unique-steps}. The DAG property is only used to show that a token never goes twice to the same vertex. This in turn bounds the number of possible reconfiguration sequences in collapses of our original DAG (\cref{lem:upper-limit-for-sequence-lengths}).

Our proof then directly provide the following result: Given an (undirected) instance $(G,S,D)$ for \ists, deciding whether there is a reconfiguration sequence where no token goes twice to the same vertex can be decided in \FPT, parameterized by $k+\tw(G)$.

To generalize this idea, we introduce another parameter.
\begin{definition}
	Given a reconfiguration sequence $(\alpha_i)_{i\in I}$, a token $t$ and a vertex $v$, the {\em iteration} of $(v,t)$ is the number of time the token enters $v$ (plus one if $t$ starts in~$v$).
	The {\em iteration}  $\iota$ of a reconfiguration sequence $(\alpha_i)_{i\in I}$ is the maximum iteration of all pair~$(v,t)$:\\	
	$\iota := \max\limits_{(v,t)\in V\times T}|\{0~|~\alpha_0(t) = v\} \cup \{i\in I~|~0<i\textrm{ and } \alpha_{i-1}(t)\neq\alpha_{i}(t) = v\}|$.
		
\end{definition}

For galactic graphs, the iteration is only defined on planets, with no constrains on the number of time a token enters a black holes. Observe then that when taking the collapse of a reconfiguration sequence, its iteration does not increase.

\begin{lemma}[Analogue of \cref{lem:upper-limit-for-sequence-lengths}]
	\label{lem:upper-limit-iota}
	Let $G$ be a (undirected) galactic graph with no neighboring black holes; the number and length of reconfiguration sequences with $k$ tokens and of iteration $\iota$ over $G$ is bounded by $g(|V(G)|, k, \iota)$ for some computable function $g$.
\end{lemma}
\begin{proof}
	First observe that with the absence of neighboring black holes, each step introduces a change for a planet (losing or receiving a token). As, a planet only witness $2k\iota$ many changes, this bounds the length of every sequence by $2k\iota n$, where $n=|V(G)|$.

	As there are $n^k$ possible configurations, the total number of possible reconfiguration sequences of iteration $\iota$ is at most $(n^k)^{2k\iota n} = n^{2k^2\iota n}$.
\end{proof}

We then call the {\em $\iota$-profile} of $x$ the set of reconfiguration~$\alpha$ of iteration $\iota$ over $\warp(x)$ such that there is a reconfiguration $\beta$ over $\significant(x)$ that collapses to $\alpha$ over $\warp(x)$.

\begin{proposition}[Annalogue of \cref{prop:dp-algo}]\label{dp-algo-iota}
	Given a tree decomposition of a graph $G$ of width \tw, a node $x$ of the decomposition, and the $\iota$-profile of its children $y_1$ and $y_2$, we can compute the $\iota$-profile of $x$ in time $f(k,\tw,\iota)$.
\end{proposition}

Which in turn yields the proof for \cref{theo:fpt-iota}.

	\section{Conclusion}
	We used galactic graphs, to show that \isdts in DAGs is tractable, when parameterized by treewidth and the number of tokens $k$. Otherwise the problem becomes hard, unless either the depth is at most 2, or at most 3 and $k$ small. Noticeable, the treewidth result applies to undirected graphs, if the iteration $\iota$ is restricted.

Thus, it might be interesting, to study this parameters behavior in graph classes other then DAGs. Interesting further results might also arise from applying the used techniques to other problems (Dominating Set Reconfiguration might be a good candidate) or to generalize to other directed graph classes that resemble DAGs.
	
	\pagebreak
    \bibliography{ref.bib}
\end{document}